\documentclass{article}
\usepackage[letterpaper, margin=0.7in]{geometry}

\usepackage[utf8]{inputenc}

\usepackage[hyphens]{url}
\usepackage[hidelinks]{hyperref}

\usepackage{booktabs}

\usepackage[inline]{enumitem}

\usepackage{tablefootnote}
\usepackage{multirow}

\usepackage{adjustbox}

\usepackage{comment}

\usepackage{graphicx}
\usepackage{framed}

\begin{document}

\title{On Automatic Parsing of Log Records}

\author{ Jared Rand and Andriy Miranskyy \\
Department of Computer Science, Ryerson University, Toronto, Canada \\
jrand@ryerson.ca, avm@ryerson.ca
}
\date{}

\maketitle

\begin{abstract}
Software log analysis helps to maintain the health of software solutions and ensure compliance and security. Existing software systems consist of heterogeneous components emitting logs in various formats. A typical solution is to unify the logs using manually built parsers, which is laborious.

Instead, we explore the possibility of automating the parsing task by employing machine translation (MT). We create a tool that generates synthetic Apache log records which we used to train recurrent-neural-network-based MT models. Models' evaluation on real-world logs shows that the models can learn Apache log format and parse individual log records. The median relative edit distance between an actual real-world log record and the MT prediction is less than or equal to 28\%. Thus, we show that log parsing using an MT approach is promising.

\end{abstract}

\section{Introduction}
Modern software solutions consist of a stack of hardware and software components~\cite{DBLP:conf/acsac/YenOOLRJK13,DBLP:journals/software/MiranskyyHCL16}. A failure of a component may lead to an outage, affecting the solution’s customers. It is crucial to quickly detect such a failure and identify the potential root cause of the problem to create a fix. Detection of the root cause is a daunting task, taking up to 40\% of the time needed to fix the problem~\cite{DBLP:journals/jss/MurtazaHMG14}. Finding the root causes is done by examining the logs emitted by the solution’s components~\cite{DBLP:journals/jss/MurtazaHMG14,DBLP:conf/sigsoft/MiranskyyMGDWG07,DBLP:conf/icse/BeschastnikhBEK14,DBLP:journals/tse/MarianiPS17}. 
Additionally, these logs may be used in the detection of cyberattacks~\cite{DBLP:conf/acsac/YenOOLRJK13} or auditing compliance with policies~\cite{DBLP:conf/acsac/YenOOLRJK13,DBLP:conf/IEEEares/DernaikaCCR19}.

There exist tools that do automatic log examinations and speed up diagnostics~\cite{DBLP:journals/jss/MurtazaHMG14,DBLP:conf/sigsoft/MiranskyyMGDWG07,DBLP:conf/icse/BeschastnikhBEK14,DBLP:journals/tse/MarianiPS17}. Unfortunately, components have individual log formats, while analysis tools prefer the logs in a unified format~\cite{DBLP:journals/software/MiranskyyHCL16}. 
Thus, maintainers need to manually create individual converters for every log format~\cite{DBLP:conf/icws/HeZZL17}. Given that a solution may have thousands of components, this becomes laborious~\cite{DBLP:journals/software/MiranskyyHCL16,DBLP:conf/iwpc/MessaoudiPBBS18}. If one can create a tool that parses logs automatically,  this tool will speed up defect detection and repair, allowing developers to focus on creating new functionality and reducing maintenance efforts. 

There exists a significant body of literature on recognizing formats. Researchers worked on detecting field type\footnote{Note that it is relatively easy to detect some patterns, e.g., dates, using basic RegEx. But a basic RegEx may have challenges understanding if a date belongs to the log record's timestamp field or if the date is part of a log message itself, which implies that developers will have to code up additional rules. } in a structured record set~\cite{DBLP:conf/kdd/HulsebosHBZSKDH19}, developing parsers rapidly and incrementally~\cite{DBLP:conf/csmr/NierstraszKGLB07}, extracting specific information from a log string~\cite{DBLP:conf/icdm/FuLWL09,DBLP:journals/tkde/MakanjuZM12,DBLP:conf/icws/HeZZL17,DBLP:conf/IEEEares/DernaikaCCR19}, and detecting log formats by comparing multiple similar log strings~\cite{DBLP:conf/icdm/Du016,DBLP:conf/iwpc/MessaoudiPBBS18,DBLP:journals/infsof/El-MasriPGHB20}. \textit{But can we individually parse a complete raw log string  (so that we can convert it to a universal format) without comparing it to other log records emitted by the same component? }

This leads us to the following research question (RQ): \textit{How can we automatically parse an individual raw log string?} To answer our RQ, we reduce parsing to a machine translation (MT) task. That is, we need to create an Oracle that, given a raw log string as input, will produce a string of tokens, showing which particular field a character in the input string belongs to. A toy example of the input and output is given in Figure~\ref{fig:toy}.

How should an Oracle look? In the field of MT for natural languages,  the best models are currently based on deep neural networks (DNNs) following an encoder-decoder architecture that implements sequence-to-sequence learning.  We will discuss the details of the models in Section~\ref{sec:models}.

DNNs require large volumes of data for training. Thus, we create a tool to generate synthetic logs mimicking real ones. We will further discuss synthetic data generation and real logs (used for validation of the models) in Section~\ref{sec:data}. 
We will then discuss the setup of our experiments in Section~\ref{sec:setup}, followed by a comparison of the performance of our models in Section~\ref{sec:results} and a summary of our findings in Section~\ref{sec:summary}. Finally, we will provide details for accessing the tool and the logs in Section~\ref{sec:access}.

\begin{figure}[bt]
    \centering
    \begin{framed}
    \small
    \texttt{2020-09-20 jane ERROR file not found} \\
    \texttt{tttttttttt\_uuuu\_lllll\_iiiiiiiiiiiiii} 
    \end{framed}
    \caption{A toy example of log translation. The top line represents an input string of the raw log. The characters in the bottom line represent the field type that a given input character is mapped to. For example, all the characters in the substring \texttt{2020-09-20} belong to a time field, denoted by \texttt{t}; substring \texttt{jane} represents a user field denoted by \texttt{u};  substring \texttt{ERROR} is a log record type field denoted by \texttt{l}, and  substring \texttt{file not found} is log message details denoted by \texttt{i}. Finally, \texttt{\_} denotes a separator between the fields. }
    \label{fig:toy}
\end{figure}

\section{Models} \label{sec:models}
We chose DNN architectures, popular for machine translation of natural language. First, we compared recurrent neural networks using gated recurrent units  (GRU)~\cite{DBLP:conf/emnlp/ChoMGBBSB14} against those using long short-term memory (LSTM)~\cite{DBLP:journals/neco/HochreiterS97} cells. Our preliminary experiments showed that LSTM-based translators outperformed the GRU-based ones (hinting that extra memory-related gates present in the LSTM cell but absent in the GRU cell may be necessary for our task). Thus, we focused on LSTM-based translators and converged on three architectures. 

The first one, deemed $M_C$, is based on the classic neural MT architecture akin to~\cite{DBLP:conf/nips/SutskeverVL14} with LSTM-based encoder and decoder layers. The second one, deemed $M_L$, is similar to $M_C$ but added the attention layer as per Bahdanau et al.~\cite{DBLP:journals/corr/BahdanauCB14}. The third one, deemed $M_S$, is based on the seq2seq architecture~\cite{DBLP:journals/corr/BritzGLL17} with the LSTM layers and  Luong et al. attention mechanism~\cite{DBLP:conf/emnlp/LuongPM15}. During our initial tests we evaluated $M_S$ with and without beam search decoding~\cite{DBLP:journals/corr/WuSCLNMKCGMKSJL16}. We ended up using $M_S$ with regular inference as initial tests found beam search evaluation to be inferior for our datasets.

\section{Logs under study} \label{sec:data}
In this work, we focus on log formats of the  Apache HTTP server~\cite{apachemodlog:online}. The ubiquitousness of this product makes it a good candidate for our tests.

We will discuss our generator tool, designed to create synthetic Apache log records in Section~\ref{sec:generator}. Then we will discuss the real-world datasets in Section~\ref{sec:real_data}, followed by the description of the synthetic datasets used for training the Oracle in Section~\ref{sec:synth_data}.  

\subsection{Synthetic logs' generator} \label{sec:generator}
As discussed above, to train DNNs, we need large volumes of data. There exist tools for generating synthetic logs, e.g.,~\cite{faker:online,kiritbas44:online}. However, for the training of the DNNs, we need not only the raw log strings, which will serve as input to a DNN, but also the information about the log strings' format that will be used as output from the model. 

Most of the generators use the Apache Common Log Format (hereby referred to as CLF) or Combined Log Format (an extended version of CLF hereby referred to as ELF). Explanations of what these formats look like can be found in Table~\ref{tab:formats} (for details, see Apache manual~\cite{apachecombinedlogformat:online}).

While CLF and ELF are the most common forms of Apache logs, we did not feel confident that a DNN trained solely on CLF and ELF could accomplish our goal of recognizing the fields of any Apache log. Additionally, the existing fake log generators did not vary much in their generated data, often using the current date of generation as the date for each log record and using a small subset of fake websites and sub-pages for which it would create requests. Furthermore, in order to train the DNNs, we needed `ground truth' output strings, which would be easier to generate alongside the log strings rather than after the fact.

Thus, we created our own Python-based generator, which extends and combines the Faker~\cite{faker:online} and fake-useragent~\cite{fakeuser:online} libraries. These libraries were used to realistically generate specifically-formatted fields, such as the user agent field and IPv6 addresses. 

Our generator creates synthetic log records based on 15 fields\footnote{There are $\approx 3.6 \times 10^{12}$ permutations of log formats that can be generated from these 15 fields. For efficiency, we only implemented these 15 fields out of 43 total possible Apache fields~\cite{apachemodlog:online}. The generator can be easily extended with additional fields, if required. } and one separator field listed in Table~\ref{tab:fields}. The user should specify the total number of the log records to produce and the type of format (random or fixed). 
The tool then generates raw log records (input) and the associated translations (outputs). An example of the generated data is shown in Figure~\ref{fig:synth_record}.

\begin{table}[bt]
\caption{ Apache log formats, see Table~\ref{tab:fields} for fields' description.}
\label{tab:formats}
\centering
\begin{tabular}{@{}ll@{}}
\toprule
Name                             & Fields (acronyms)       \\ 
\midrule
Common Log Format (CLF)          & \texttt{h l u t "r" s b}         \\
Combined Log Format (ELF) & \texttt{h l u t "r" s b "R" "i"} \\ 
\bottomrule
\end{tabular}
\end{table}

\begin{table}[bt]
\caption{A list of fields that our synthetic log generator can produce. For details about the fields, see~\cite{apachemodlog:online}.}
\label{tab:fields}
\centering
\begin{tabular}{@{}p{0.09\columnwidth}p{0.87\columnwidth}@{}}
\toprule
Field & Field           \\                                                         
acronym & description   \\ \midrule
\texttt{h}             & IP address of the client host. Can be IPv4 or IPv6. \\
\texttt{l}             & The remote logname. We were unable to find a good example of what   kinds of values are returned by Apache servers, thus for this  paper we only supplied the commonly-given value ‘-’. This field could not be omitted as it is present in both the ELF and CLF formats.  \\
\texttt{u}             & The remote username. Can be empty (‘-’) \\
\texttt{t}             & The datetime of the request, presented in the default {[}day/month/year:hour:minute:second zone{]} format.  \\
\texttt{r}             & The request line from the client. Made up of the method, path and   querystring, and protocol. \\
\texttt{s}             & The status of the request. \\
\texttt{b}             & The number of bytes sent.   \\
\texttt{m}             & The request method.   \\
\texttt{U}             & The requested URI path.  \\
\texttt{H}             & The request protocol.  \\
\texttt{q}             & The request querystring.   \\
\texttt{v}             & The canonical servername of the server servicing the request.  \\
\texttt{V}             & The servername according to UseCanonical. In our generator, this   field is identical to the \texttt{v}   field.   \\
\texttt{i}            & The user agent of the request\tablefootnote{\label{extrafieldnote} In a real Apache HTTP server deployment, this field is extracted from the \texttt{\%i}  log parameter, see~\cite{apachemodlog:online} for details.}. \\
\texttt{R}             & The referrer of the request\textsuperscript{\ref{extrafieldnote}}. \\   
\texttt{\_}             & Represents a separator between log fields. \\
\bottomrule
\end{tabular}
\end{table}

\begin{figure*}[t]
    \centering
    \tiny
    \begin{framed}
    \texttt{1: 41.193.93.229 - - 10/Jul/7983:05:08:49 +0100 "GET explore/category/home.html HTTP/2" 302 65953} \\
    \texttt{2: hhhhhhhhhhhhh\_l\_u\_tttttttttttttttttttttttttt\_rrrrrrrrrrrrrrrrrrrrrrrrrrrrrrrrrrrrrrr\_sss\_bbbbb} \\
    \texttt{3: 192.168.4.25 - - [22/Dec/2016:16:11:41 +0300] "POST /DVWA/login.php HTTP/1.1" 200 1532 "-" "Mozilla/4.0 (compatible; MSIE 8.0; Windows NT 6.1; Trident/4.0; w3af.sf.net"} \\
    \texttt{4: hhhhhhhhhhhh\_l\_u\_[tttttttttttttttttttttttttt]\_"rrrrrrrrrrrrrrrrrrrrrrrrrrrrr"\_sss\_bbbb\_"R"\_"iiiiiiiiiiiiiiiiiiiiiiiiiiiiiiiiiiiiiiiiiiiiiiiiiiiiiiiiiiiiiiiiiiiiiiiiiii"} \\
    \texttt{5: PUT ycgmvjc.com HTTP/2 - 102186 - \url{d8idf~gyck.html}} \\
    \texttt{6: mmm\_VVVVVVVVVVV\_HHHHHH\_l\_bbbbbb\_F\_UUUUUUUUUUUUUUU}
    \end{framed}
    \caption{An example of a record for translation generated by our log generator.  Lines 1, 3, and 5 are examples of logs in CLF, ELF, and random format, respectively. The characters in lines 2, 4, and 6 represent mapping to their equivalent field type (described in Table~\ref{tab:fields}).  }
    \label{fig:synth_record}
\end{figure*}

\subsection{Real logs} \label{sec:real_data}

To validate the generalizability of the DNNs trained on synthetic log records produced by the generator discussed in Section~\ref{sec:generator}, we take three publicly available log files and deem the validation datasets $V_A$~\cite{apacheht89:online}, $V_B$~\cite{apacheht41:online}, and $V_C$~\cite{examples55:online}. The first two logs capture execution results of two web vulnerability scanners, namely, Netsparker and Acunetix~\cite{BASSEYYAR201828}. The third one is a sample Apache web server log file that can be processed by Elastic software. The summary statistics of the log files is shown in Table~\ref{tab:datasets}.

The log format of the validation logs is (or is extremely similar to) ELF\footnote{The log records of $V_A$ and $V_B$ were wrapped in quotes (\texttt{"}), which makes them a slightly different format from ELF. While this is likely a product of processing and these quotes could be removed, we opted to keep them for training completeness.}. We have parsed the log records from these three files using custom scripts to produce the correct character-to-field mapping strings. 

\subsection{Synthetic datasets} \label{sec:synth_data}

Using our log generator (covered in Section~\ref{sec:generator}), we create synthetic logs for training and testing, while varying the number of log records and their formats. We create five synthetic datasets that may help us uncover various challenges of the ``log translation'' task. The datasets summary statistics are given in Table~\ref{tab:datasets}; their details are as follows.

The first dataset, deemed $T_T$, contains \textit{trivial}-to-learn data. 
The triviality of this dataset refers to the ability of models trained on this set to learn the fields and correctly interpret the validation datasets. $T_T$ consists of 100K ELF mock logs. The models trained on this dataset should be able to recognize the ELF-based log records from our validation datasets very well. However, the generalizability to other formats would be mediocre.

The second dataset, deemed $T_E$, has \textit{easy}-to-learn data. 
It contains 20K lines using a mix of the formats: ELF, CLF, and randomly generated records, which are generated based on all the fields in Table~\ref{tab:fields}. The random and CLF records are added to assess the generalizability of the model. Hypothetically, models trained on $T_E$ and validated on $V_{(\cdot)}$ datasets will have worse performance than those trained on $T_T$, because it has fewer log records and the model has to learn of multiple formats.

The third dataset, deemed $T_M$, contains \textit{moderately}-difficult-to-learn data.
$T_M$ contains a mix of formats: $\approx 50\%$ of 100K records are in the CLF format and $\approx 50\%$ have randomly generated formats. Based on  Table~\ref{tab:formats},  CLF is similar to ELF, but two of the fields (\texttt{R} and \texttt{i}) are missing. Thus, it will be harder for the model to recognize ELF records in $V_{(\cdot)}$. 

The fourth dataset, deemed $T_M^\prime$, is a shorter version of $T_M$ containing only 20K records. The more observations a dataset has, the more information a machine learning model will obtain for training. However, this comes at a cost of increased training time. Thus, we want to see how the change in observations affects models' performance.
 $T_M^\prime$ will help us assess how much data are needed for training. 

The fifth dataset, deemed $T_H$, carries \textit{hard}-to-learn data. 
This dataset contains 100K randomly generated records, and none of them look like CLF or ELF. Thus, models trained on this dataset must learn the nature of the fields well in order to properly parse log records in $V_{(\cdot)}$.

\begin{table*}[bt]
\caption{ Datasets' description.}
\label{tab:datasets}
\centering
\begin{tabular}{@{}lrrrrp{0.53\textwidth}@{}}
\toprule
Dataset & Log records&  \multicolumn{3}{c}{Log records length}   & Log records' format description \\
  \cmidrule{3-5}
 & count & min & median & max  &   \\
\midrule
$T_T$  & 100,000 & 136 & 272 & 1173              & $100\%$ ELF  \\
$T_E$  & 20,000 & 4 & 250 & 1173              & $\approx 40\%$ of the ELF format, $\approx 24\%$ of the CLF format, and $\approx 36\%$ of randomly drawn and reshuffled fields shown in Table~\ref{tab:fields}.  The random strings have $2$ to $14$ records in them.  \\
$T_M$  & 100,000 & 4 & 161 & 1294              & $\approx 50\%$ of the CLF format and $\approx 50\%$  of the randomly generated records using the same approach as in the $T_E$ case. The random strings have $2$ to $14$ fields in them.    \\
$T_M^\prime$  & 20,000 & 4 & 162 & 1527              &  Identical to the one of $T_M$.  \\
$T_H$  & 100,000 & 4 & 291 & 1528              & $100\%$ of the randomly generated records using the same approach as in the $T_E$ case. The random strings have $2$ to $15$ fields in them.    \\
\midrule
$V_A$  & 7,314 & 194 & 238 & 602 & 100\% ELF                                \\
$V_B$  & 6,539 & 79 & 238 & 4398 & 100\% ELF                                \\
$V_C$  & 10,000 & 81 & 231 & 1363  & 100\% ELF                                \\ 
\bottomrule
\end{tabular}
\end{table*}

\section{Experimental Setup} \label{sec:setup}
We want to assess the generalizability of our approach. To do this, we create three groups of experiments where a model is trained on Training Datasets $T_{(\cdot)}$ (described in Section~\ref{sec:synth_data}). We train the neural networks for up to 300 epochs with a mini-batch size of 64 log records. Optimization is done using the Adam algorithm\footnote{We used learning rate of $0.001$, $\beta_1=0.9$, $\beta_2=0.999$, and $\epsilon=10^{-7}$.}~\cite{DBLP:journals/corr/KingmaB14}. Each dataset is randomly split into 90\% training data and 10\% validation data during training.  The best model, preserved for further evaluation, is the one with the lowest cross-entropy loss for $M_C$ and $M_L$ and sparse categorical cross-entropy loss for $M_S$. Translation is done at the character-level, i.e., we do not do any log record prepossessing.

We have tuned the models' hyper-parameters. For $M_C$ and $M_L$, we have experimented with different number of cells in these layers, namely $\{256, 512, 1024\}$. This was done to assess the degree of `freedom' required to recognize different log formats. We also varied the dropout rates in the encoder and decoders layer, setting them to $\{0.0, 0.2, 0.4, 0.6, 0.8\}$. This was done to regularize the model and reduce overtraining.

We evaluate the performance of the models on three real-world log files (discussed in Section~\ref{sec:real_data}). The performance is measured using the Levenshtein (a.k.a. the edit) distance~\cite{levenshtein1966binary}, which computes the number of changes needed to transform a string returned by the Oracle into an expected (ground truth) string. 
We report the absolute and relative Levenshtein distances, deemed $D_A$ and $D_R$, respectively. For a given log record, $D_R$ is computed by dividing $D_A$ by the log record length. $D_A$ and $D_R$ lie within $[0, \infty)$ range: the closer to $0$~--- the better. 

\section{Results} \label{sec:results}

\subsection{Comparison approach}
\begin{figure}[bt]
    \centering
    \includegraphics[width=\columnwidth]{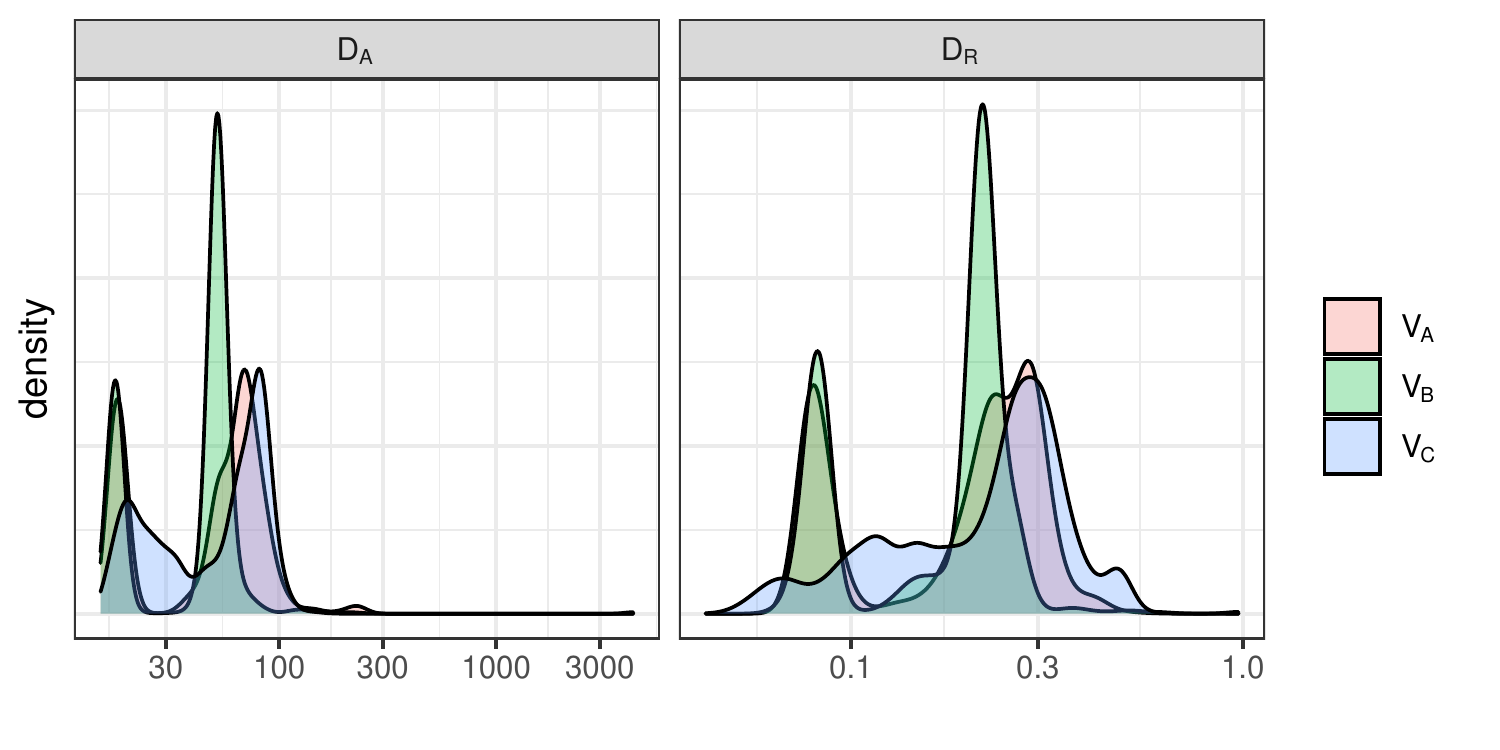}
    \caption{Distribution of $D_A$ and $D_R$ values (left and right pane, respectively) for the best model trained on the $T_M$ dataset and validated on $V_A$, $V_B$, and $V_C$ datasets.}
    \label{fig:distr}
\end{figure}

It is easy to measure $D_A$ and $D_R$ for individual strings. However, how to aggregate the measurements for a particular $V_{(\cdot)}$ dataset? Let us look at the distributions of the $D_A$ and $D_R$ values. For the sake of brevity, we show the distributions only for the models trained on $T_M$ in Figure~\ref{fig:distr}. As we can see, the distributions have a long right tail: although the median values of $D_A$ are between 51 and 61, the max values go up to 4283. This is due to the presence in $V_B$ and $V_C$ of a small number of long log records (as shown in Table~\ref{tab:datasets}). Namely, less than  0.2\% of log records are longer than 1000 characters, while the median length of the log records is between 231 and 238 characters for all three $V_{(\cdot)}$. 
Once we normalize the results using the  $D_R$, we can see that the amount of changes is proportional to the string’s length, with the median $D_R$ between $\approx$~21 and 25\%.

To preserve space and allow quantitative comparison, we present the summary statistics for the remaining evaluations by computing min, max, mean, and various quantiles of the distributions. The results for the top models, based on the lowest values of $D_A$ and $D_R$ for a given quantile, are shown in Table~\ref{tab:results}.

\setlength{\tabcolsep}{2.6pt}
\begin{table*}[tb]
\caption{ Summary statistics for the distributions of $D_A$ and $D_R$ for the best performing models. All the models have 512 cells in the LSTM layers; $d$ denotes the dropout rate and is either $0.2$ or $0.4$. $q_p$ denotes the sample quantile function, where $p$ is the probability value; e.g., when $p=0.50$, the $q_{.50}$ returns the median of a distribution. For example, for a model trained on dataset $T_H$, when validated on $V_C$, the maximum value of $D_R$ is $0.91$ (as shown in the bottom-right corner). }
\label{tab:results}
\begin{adjustbox}{max width=\textwidth}
\begin{tabular}{@{}ll|rrrrrrrr|rrrrrrrr|rrrrrrrr@{}}
\toprule
 \multicolumn{2}{c|}{Trained}     & \multicolumn{8}{c|}{$V_A$}                                                 & \multicolumn{8}{c|}{$V_B$}                                                 & \multicolumn{8}{c}{$V_C$}                                                 \\
  \multicolumn{2}{c|}{on}            & min  & avg  & $q_{.50}$ & $q_{.75}$ & $q_{.90}$ & $q_{.95}$ & $q_{.99}$ & max  & min  & avg  & $q_{.50}$ & $q_{.75}$ & $q_{.90}$ & $q_{.95}$ & $q_{.99}$ & max  & min  & avg  & $q_{.50}$ & $q_{.75}$ & $q_{.90}$ & $q_{.95}$ & $q_{.99}$ & max  \\
\midrule
\multirow{6}{*}{$D_A$} & $T_T, d = 0.2$ & 7 & 33 & 13 & 21 & 101 & 119 & 224 & 298 & 7 & 22 & 14 & 14 & 17 & 33 & 105 & 4150 & 2 & 22 & 9 & 17 & 78 & 83 & 169 & 419 \\
& $T_E, d = 0.2$        & 13 & 43 & 22 & 32 & 121 & 151 & 277 & 426 & 11 & 30 & 23 & 23 & 25 & 51 & 140 & 3566 & 7  & 23 & 18 & 24 & 33  & 59  & 77  & 1050 \\
& $T_E, d = 0.4$        & 10 & 35 & 20 & 25 & 59  & 122 & 305 & 491 & 10 & 29 & 19 & 20 & 22 & 45 & 136 & 4145 & 6  & 22 & 13 & 21 & 37  & 102 & 174 & 853  \\
& $T_M^\prime, d = 0.2$ & 21 & 63 & 56 & 78 & 100 & 122 & 305 & 477 & 22 & 61 & 58 & 58 & 65 & 78 & 204 & 4242 & 17 & 70 & 70 & 91 & 113 & 116 & 184 & 1281 \\
& $T_M, d = 0.2$        & 17 & 55 & 59 & 71 & 86  & 92  & 203 & 240 & 16 & 51 & 51 & 53 & 57 & 59 & 124 & 4283 & 15 & 56 & 61 & 80 & 88  & 93  & 108 & 1146 \\
& $T_H, d = 0.2$        & 15 & 69 & 68 & 88 & 113 & 135 & 233 & 476 & 16 & 60 & 59 & 61 & 67 & 97 & 201 & 4285 & 15 & 66 & 66 & 85 & 105 & 110 & 202 & 1241 \\
\midrule
\multirow{6}{*}{$D_R$} & $T_T, d = 0.2$            & 0.03 & 0.11 & 0.06 & 0.08 & 0.33 & 0.36 & 0.50 & 0.59 & 0.03 & 0.06 & 0.06 & 0.06 & 0.06 & 0.11 & 0.34 & 2.18 & 0.01 & 0.12 & 0.04 & 0.09 & 0.27 & 0.50 & 1.75 & 2.02 \\
& $T_E, d = 0.2$            & 0.05 & 0.15 & 0.08     & 0.10     & 0.39     & 0.46     & 0.61     & 0.71 & 0.05 & 0.10 & 0.10     & 0.10     & 0.10     & 0.17     & 0.43     & 0.82 & 0.03 & 0.10 & 0.08     & 0.09     & 0.15     & 0.35     & 0.47     & 0.77 \\
& $T_E, d = 0.4$            & 0.05 & 0.12 & 0.08     & 0.09     & 0.20     & 0.40     & 0.69     & 0.82 & 0.04 & 0.09 & 0.08     & 0.08     & 0.10     & 0.15     & 0.46     & 0.94 & 0.03 & 0.11 & 0.05     & 0.10     & 0.20     & 0.51     & 0.63     & 0.75 \\
& $T_M^\prime, d = 0.2$           & 0.09 & 0.23 & 0.23     & 0.26     & 0.33     & 0.39     & 0.70     & 0.80 & 0.09 & 0.22 & 0.24     & 0.24     & 0.29     & 0.29     & 0.61     & 0.97 & 0.09 & 0.30 & 0.29     & 0.36     & 0.43     & 0.59     & 0.69     & 0.94 \\
& $T_M, d = 0.2$         & 0.07 & 0.20 & 0.23     & 0.28     & 0.30     & 0.33     & 0.42     & 0.58 & 0.06 & 0.18 & 0.21     & 0.22     & 0.25     & 0.26     & 0.37     & 0.97 & 0.04 & 0.24 & 0.25     & 0.31     & 0.36     & 0.42     & 0.48     & 0.84 \\
& $T_H, d = 0.2$ & 0.07 & 0.25 & 0.28 & 0.31 & 0.40 & 0.45 & 0.63 & 0.79 & 0.06 & 0.21 & 0.25 & 0.25 & 0.26 & 0.35 & 0.61 & 0.97 & 0.06 & 0.28 & 0.27 & 0.32 & 0.43 & 0.63 & 0.64 & 0.91 \\
\bottomrule
\end{tabular}
\end{adjustbox}
\end{table*}

\subsection{Discussion}

\subsubsection{Architecture comparison}
The models with the attention mechanisms ($M_L$ and $M_S$) underperformed and are not shown in Table~\ref{tab:results}; the $M_C$-based models are the winners. The poor performance may suggest that the attention mechanisms, which work well for natural language translation, are less suited for our problem.

\subsubsection{Hyperparameters}
Table~\ref{tab:results} shows that the models with 512 cells prevailed, which may imply that we do need some degree of freedom to learn the patterns but not too much to overwhelm the optimizer.

The dropout rate $d = 0.2$ is the best in all cases, except for models trained on $T_E$. In this particular case, a dropout of $0.2$ and $0.4$ provide similar results. We expected the dropout to be positive as our goal is to generalize the learning process, and these expectations were met.

\subsubsection{Dataset size}
Comparison of results for $T_M$ and $T_M^\prime$ in Table~\ref{tab:results} suggests that the larger number of observations improves performances of the model. However,  we can observe the diminishing returns: increasing the dataset size by a factor of five (as in the case of $T_M$ and $T_M^\prime$) leads to incremental improvements in the $D_A$ and $D_R$ values. 

Moreover, compare the results for models trained on $T_E$ and $T_M$: the model trained on 20K of simpler observations from $T_E$ performs better than the model trained on 100K harder records from $T_M$. This implies that the quality of the input matters more than the quantity.

\subsubsection{Training time}
We used Nvidia Tesla P100 and Titan X Pascal GPUs for the training. The training time per epoch (for datasets with 100K log records) ranged between one and three hours, depending on the models and inputs. Thus, training the model for hundreds of epochs is time-consuming. We observed that the validation loss would often reach the lowest (best) values between 60 and 150 epochs during models' training, which enabled us to perform early stopping and save time.

\subsubsection{Training datasets vs. models' performance} As expected, the resulting models' performance degrades as the difficulty of the training dataset increases. The $T_T$-based model came up first (with the exception of a few high quantiles): it had to learn only the fields' beginning and end while the fields' order stayed the same. The $T_H$-based model came last: not only did the model have to understand the beginning and ends of the fields, but also it had to learn the field's order while never been able to see a single example of the ELF record. To our pleasant surprise, the $T_H$-based model did not fall behind their competitors significantly and was quite close to the $ T_M $-based model, with the median $ D_R$ between $0.25$ and  $0.28$, and the 90-th quantile of $D_R$ between $0.26$ and  $0.43$. This implies that the $ T_H $-based model was able to learn the concept of fields relatively well.

\section{Summary} \label{sec:summary}
Our RQ was: \textit{How can we automatically parse an individual raw log string?} To answer the RQ, we reformulated parsing as an MT task and explored three MT architectures. We showed that the models based on the classic LSTM MT architecture are the most promising. 
Moreover, the model trained on randomly generated log records was able to partially parse the log records in a format that it did not see in training: 50\% of log records per evaluation dataset were parsed with the $D_R \leq 28\%$ and 90\% of records with the $D_R \leq 43\%$.  This implies that parsing of the individual strings using MT is possible.

This work is of interest to researchers as it provides novel insights into the log parsing field. The quality of our models’ translation is not adequate for the practitioner at this stage, but we hope that this work will inspire others to explore various MT approaches and improve upon our results. 

To aid these researchers, we created a tool for generating synthetic logs in Apache format used for creating training data for the models. We also found and parsed three real-world validation logs for models’ evaluation. 

\section{Data Availability}\label{sec:access}
The tool source code repository is~\cite{dat:github}. The version of the tool used in this paper is stored at~\cite{code:zenodo}. The training and validation datasets, $T_{(\cdot)}$ and $V_{(\cdot)}$, are available at~\cite{dat:zenodo}.
\bibliographystyle{IEEEtran}
\bibliography{references} 

\begin{thebibliography}{10}
\providecommand{\url}[1]{#1}
\csname url@samestyle\endcsname
\providecommand{\newblock}{\relax}
\providecommand{\bibinfo}[2]{#2}
\providecommand{\BIBentrySTDinterwordspacing}{\spaceskip=0pt\relax}
\providecommand{\BIBentryALTinterwordstretchfactor}{4}
\providecommand{\BIBentryALTinterwordspacing}{\spaceskip=\fontdimen2\font plus
\BIBentryALTinterwordstretchfactor\fontdimen3\font minus
  \fontdimen4\font\relax}
\providecommand{\BIBforeignlanguage}[2]{{%
\expandafter\ifx\csname l@#1\endcsname\relax
\typeout{** WARNING: IEEEtran.bst: No hyphenation pattern has been}%
\typeout{** loaded for the language `#1'. Using the pattern for}%
\typeout{** the default language instead.}%
\else
\language=\csname l@#1\endcsname
\fi
#2}}
\providecommand{\BIBdecl}{\relax}
\BIBdecl

\bibitem{DBLP:conf/acsac/YenOOLRJK13}
\BIBentryALTinterwordspacing
T.~Yen, A.~Oprea, K.~Onarlioglu, T.~Leetham, W.~K. Robertson, A.~Juels, and
  E.~Kirda, ``Beehive: large-scale log analysis for detecting suspicious
  activity in enterprise networks,'' in \emph{Annual Computer Security
  Applications Conference, {ACSAC} '13}.\hskip 1em plus 0.5em minus 0.4em\relax
  {ACM}, 2013, pp. 199--208. [Online]. Available:
  \url{https://doi.org/10.1145/2523649.2523670}
\BIBentrySTDinterwordspacing

\bibitem{DBLP:journals/software/MiranskyyHCL16}
\BIBentryALTinterwordspacing
A.~V. Miranskyy, A.~Hamou{-}Lhadj, E.~Cialini, and A.~Larsson,
  ``Operational-log analysis for big data systems: Challenges and solutions,''
  \emph{{IEEE} Software}, vol.~33, no.~2, pp. 52--59, 2016. [Online].
  Available: \url{https://doi.org/10.1109/MS.2016.33}
\BIBentrySTDinterwordspacing

\bibitem{DBLP:journals/jss/MurtazaHMG14}
\BIBentryALTinterwordspacing
S.~S. Murtaza, A.~Hamou{-}Lhadj, N.~H. Madhavji, and M.~Gittens, ``An empirical
  study on the use of mutant traces for diagnosis of faults in deployed
  systems,'' \emph{Journal of Systems and Software}, vol.~90, pp. 29--44, 2014.
  [Online]. Available: \url{https://doi.org/10.1016/j.jss.2013.11.1094}
\BIBentrySTDinterwordspacing

\bibitem{DBLP:conf/sigsoft/MiranskyyMGDWG07}
\BIBentryALTinterwordspacing
A.~V. Miranskyy, N.~H. Madhavji, M.~Gittens, M.~Davison, M.~Wilding, and
  D.~Godwin, ``An iterative, multi-level, and scalable approach to comparing
  execution traces,'' in \emph{Proceedings of the 6th joint meeting of the
  European Software Engineering Conference and the {ACM} {SIGSOFT}
  International Symposium on Foundations of Software Engineering, 2007}.\hskip
  1em plus 0.5em minus 0.4em\relax {ACM}, 2007, pp. 537--540. [Online].
  Available: \url{https://doi.org/10.1145/1287624.1287704}
\BIBentrySTDinterwordspacing

\bibitem{DBLP:conf/icse/BeschastnikhBEK14}
\BIBentryALTinterwordspacing
I.~Beschastnikh, Y.~Brun, M.~D. Ernst, and A.~Krishnamurthy, ``Inferring models
  of concurrent systems from logs of their behavior with csight,'' in
  \emph{36th International Conference on Software Engineering, {ICSE}
  '14}.\hskip 1em plus 0.5em minus 0.4em\relax {ACM}, 2014, pp. 468--479.
  [Online]. Available: \url{https://doi.org/10.1145/2568225.2568246}
\BIBentrySTDinterwordspacing

\bibitem{DBLP:journals/tse/MarianiPS17}
\BIBentryALTinterwordspacing
L.~Mariani, M.~Pezz{\`{e}}, and M.~Santoro, ``Gk-tail+ an efficient approach to
  learn software models,'' \emph{{IEEE} Transactions on Software Engineering},
  vol.~43, no.~8, pp. 715--738, 2017. [Online]. Available:
  \url{https://doi.org/10.1109/TSE.2016.2623623}
\BIBentrySTDinterwordspacing

\bibitem{DBLP:conf/IEEEares/DernaikaCCR19}
\BIBentryALTinterwordspacing
F.~Dernaika, N.~Cuppens{-}Boulahia, F.~Cuppens, and O.~Raynaud, ``Semantic
  mediation for {A} posteriori log analysis,'' in \emph{Proceedings of the 14th
  International Conference on Availability, Reliability and Security, {ARES}
  2019}.\hskip 1em plus 0.5em minus 0.4em\relax {ACM}, 2019, pp. 88:1--88:10.
  [Online]. Available: \url{https://doi.org/10.1145/3339252.3340104}
\BIBentrySTDinterwordspacing

\bibitem{DBLP:conf/icws/HeZZL17}
\BIBentryALTinterwordspacing
P.~He, J.~Zhu, Z.~Zheng, and M.~R. Lyu, ``Drain: An online log parsing approach
  with fixed depth tree,'' in \emph{2017 {IEEE} International Conference on Web
  Services, {ICWS} 2017}.\hskip 1em plus 0.5em minus 0.4em\relax {IEEE}, 2017,
  pp. 33--40. [Online]. Available: \url{https://doi.org/10.1109/ICWS.2017.13}
\BIBentrySTDinterwordspacing

\bibitem{DBLP:conf/iwpc/MessaoudiPBBS18}
\BIBentryALTinterwordspacing
S.~Messaoudi, A.~Panichella, D.~Bianculli, L.~C. Briand, and R.~Sasnauskas, ``A
  search-based approach for accurate identification of log message formats,''
  in \emph{Proceedings of the 26th Conference on Program Comprehension, {ICPC}
  2018}.\hskip 1em plus 0.5em minus 0.4em\relax {ACM}, 2018, pp. 167--177.
  [Online]. Available: \url{https://doi.org/10.1145/3196321.3196340}
\BIBentrySTDinterwordspacing

\bibitem{DBLP:conf/kdd/HulsebosHBZSKDH19}
\BIBentryALTinterwordspacing
M.~Hulsebos \emph{et~al.}, ``Sherlock: {A} deep learning approach to semantic
  data type detection,'' in \emph{Proceedings of the 25th {ACM} {SIGKDD}
  International Conference on Knowledge Discovery {\&} Data Mining, {KDD}
  2019}.\hskip 1em plus 0.5em minus 0.4em\relax {ACM}, 2019, pp. 1500--1508.
  [Online]. Available: \url{https://doi.org/10.1145/3292500.3330993}
\BIBentrySTDinterwordspacing

\bibitem{DBLP:conf/csmr/NierstraszKGLB07}
\BIBentryALTinterwordspacing
O.~Nierstrasz, M.~Kobel, T.~G{\^{\i}}rba, M.~Lanza, and H.~Bunke,
  ``Example-driven reconstruction of software models,'' in \emph{11th European
  Conference on Software Maintenance and Reengineering, Software Evolution in
  Complex Software Intensive Systems, {CSMR} 2007}.\hskip 1em plus 0.5em minus
  0.4em\relax {IEEE} Computer Society, 2007, pp. 275--286. [Online]. Available:
  \url{https://doi.org/10.1109/CSMR.2007.23}
\BIBentrySTDinterwordspacing

\bibitem{DBLP:conf/icdm/FuLWL09}
\BIBentryALTinterwordspacing
Q.~Fu, J.~Lou, Y.~Wang, and J.~Li, ``Execution anomaly detection in distributed
  systems through unstructured log analysis,'' in \emph{{ICDM} 2009, The Ninth
  {IEEE} International Conference on Data Mining}.\hskip 1em plus 0.5em minus
  0.4em\relax {IEEE} Computer Society, 2009, pp. 149--158. [Online]. Available:
  \url{https://doi.org/10.1109/ICDM.2009.60}
\BIBentrySTDinterwordspacing

\bibitem{DBLP:journals/tkde/MakanjuZM12}
\BIBentryALTinterwordspacing
A.~Makanju, A.~N. Zincir{-}Heywood, and E.~E. Milios, ``A lightweight algorithm
  for message type extraction in system application logs,'' \emph{{IEEE}
  Transactions on Knowledge and Data Engineering}, vol.~24, no.~11, pp.
  1921--1936, 2012. [Online]. Available:
  \url{https://doi.org/10.1109/TKDE.2011.138}
\BIBentrySTDinterwordspacing

\bibitem{DBLP:conf/icdm/Du016}
\BIBentryALTinterwordspacing
M.~Du and F.~Li, ``Spell: Streaming parsing of system event logs,'' in
  \emph{{IEEE} 16th International Conference on Data Mining, {ICDM}
  2016}.\hskip 1em plus 0.5em minus 0.4em\relax {IEEE} Computer Society, 2016,
  pp. 859--864. [Online]. Available:
  \url{https://doi.org/10.1109/ICDM.2016.0103}
\BIBentrySTDinterwordspacing

\bibitem{DBLP:journals/infsof/El-MasriPGHB20}
\BIBentryALTinterwordspacing
D.~El{-}Masri, F.~Petrillo, Y.~Gu{\'{e}}h{\'{e}}neuc, A.~Hamou{-}Lhadj, and
  A.~Bouziane, ``A systematic literature review on automated log abstraction
  techniques,'' \emph{Information and Software Technology}, vol. 122, p.
  106276, 2020. [Online]. Available:
  \url{https://doi.org/10.1016/j.infsof.2020.106276}
\BIBentrySTDinterwordspacing

\bibitem{DBLP:conf/emnlp/ChoMGBBSB14}
\BIBentryALTinterwordspacing
K.~Cho, B.~van Merrienboer, {\c{C}}.~G{\"{u}}l{\c{c}}ehre, D.~Bahdanau,
  F.~Bougares, H.~Schwenk, and Y.~Bengio, ``Learning phrase representations
  using {RNN} encoder-decoder for statistical machine translation,'' in
  \emph{Proceedings of the 2014 Conference on Empirical Methods in Natural
  Language Processing, {EMNLP} 2014, {A} meeting of SIGDAT, a Special Interest
  Group of the {ACL}}.\hskip 1em plus 0.5em minus 0.4em\relax {ACL}, 2014, pp.
  1724--1734. [Online]. Available: \url{https://doi.org/10.3115/v1/d14-1179}
\BIBentrySTDinterwordspacing

\bibitem{DBLP:journals/neco/HochreiterS97}
\BIBentryALTinterwordspacing
S.~Hochreiter and J.~Schmidhuber, ``Long short-term memory,'' \emph{Neural
  Computation}, vol.~9, no.~8, pp. 1735--1780, 1997. [Online]. Available:
  \url{https://doi.org/10.1162/neco.1997.9.8.1735}
\BIBentrySTDinterwordspacing

\bibitem{DBLP:conf/nips/SutskeverVL14}
\BIBentryALTinterwordspacing
I.~Sutskever, O.~Vinyals, and Q.~V. Le, ``Sequence to sequence learning with
  neural networks,'' in \emph{Advances in Neural Information Processing Systems
  27: Annual Conference on Neural Information Processing Systems 2014}, 2014,
  pp. 3104--3112. [Online]. Available:
  \url{http://papers.nips.cc/paper/5346-sequence-to-sequence-learning-with-neural-networks}
\BIBentrySTDinterwordspacing

\bibitem{DBLP:journals/corr/BahdanauCB14}
\BIBentryALTinterwordspacing
D.~Bahdanau, K.~Cho, and Y.~Bengio, ``Neural machine translation by jointly
  learning to align and translate,'' in \emph{3rd International Conference on
  Learning Representations, {ICLR} 2015}, 2015. [Online]. Available:
  \url{http://arxiv.org/abs/1409.0473}
\BIBentrySTDinterwordspacing

\bibitem{DBLP:journals/corr/BritzGLL17}
\BIBentryALTinterwordspacing
D.~Britz, A.~Goldie, M.~Luong, and Q.~V. Le, ``Massive exploration of neural
  machine translation architectures,'' \emph{CoRR}, vol. abs/1703.03906, 2017.
  [Online]. Available: \url{http://arxiv.org/abs/1703.03906}
\BIBentrySTDinterwordspacing

\bibitem{DBLP:conf/emnlp/LuongPM15}
\BIBentryALTinterwordspacing
T.~Luong, H.~Pham, and C.~D. Manning, ``Effective approaches to attention-based
  neural machine translation,'' in \emph{Proceedings of the 2015 Conference on
  Empirical Methods in Natural Language Processing, {EMNLP} 2015}.\hskip 1em
  plus 0.5em minus 0.4em\relax The Association for Computational Linguistics,
  2015, pp. 1412--1421. [Online]. Available:
  \url{https://doi.org/10.18653/v1/d15-1166}
\BIBentrySTDinterwordspacing

\bibitem{DBLP:journals/corr/WuSCLNMKCGMKSJL16}
\BIBentryALTinterwordspacing
Y.~Wu \emph{et~al.}, ``Google's neural machine translation system: Bridging the
  gap between human and machine translation,'' \emph{CoRR}, vol.
  abs/1609.08144, 2016. [Online]. Available:
  \url{http://arxiv.org/abs/1609.08144}
\BIBentrySTDinterwordspacing

\bibitem{apachemodlog:online}
\BIBentryALTinterwordspacing
Apache module mod\_log\_config. [Online]. Available:
  \url{https://httpd.apache.org/docs/current/mod/mod_log_config.html}
\BIBentrySTDinterwordspacing

\bibitem{faker:online}
\BIBentryALTinterwordspacing
Faker package documentation. [Online]. Available:
  \url{https://faker.readthedocs.io/en/master/#}
\BIBentrySTDinterwordspacing

\bibitem{kiritbas44:online}
\BIBentryALTinterwordspacing
kiritbasu/fake-apache-log-generator: Generate a boatload of fake apache log
  files very quickly. [Online]. Available:
  \url{https://github.com/kiritbasu/Fake-Apache-Log-Generator}
\BIBentrySTDinterwordspacing

\bibitem{apachecombinedlogformat:online}
\BIBentryALTinterwordspacing
Log files. [Online]. Available:
  \url{https://httpd.apache.org/docs/1.3/logs.html#combined}
\BIBentrySTDinterwordspacing

\bibitem{fakeuser:online}
\BIBentryALTinterwordspacing
fake-useragent package documentation. [Online]. Available:
  \url{https://fake-useragent.readthedocs.io/en/stable/}
\BIBentrySTDinterwordspacing

\bibitem{apacheht89:online}
\BIBentryALTinterwordspacing
apache-http-logs/netsparker.txt at master · ocatak/apache-http-logs. [Online].
  Available:
  \url{https://github.com/ocatak/apache-http-logs/blob/master/netsparker.txt}
\BIBentrySTDinterwordspacing

\bibitem{apacheht41:online}
\BIBentryALTinterwordspacing
apache-http-logs/acunetix.txt at master · ocatak/apache-http-logs. [Online].
  Available:
  \url{https://github.com/ocatak/apache-http-logs/blob/master/acunetix.txt}
\BIBentrySTDinterwordspacing

\bibitem{examples55:online}
\BIBentryALTinterwordspacing
examples/apache\_logs at master · elastic/examples. [Online]. Available:
  \url{https://github.com/elastic/examples/blob/master/Common%20Data%20Formats/apache_logs/apache_logs}
\BIBentrySTDinterwordspacing

\bibitem{BASSEYYAR201828}
\BIBentryALTinterwordspacing
M.~B. Seyyar, F.~Özgür Çatak, and E.~Gül, ``Detection of attack-targeted
  scans from the apache http server access logs,'' \emph{Applied Computing and
  Informatics}, vol.~14, no.~1, pp. 28--36, 2018. [Online]. Available:
  \url{https://doi.org/10.1016/j.aci.2017.04.002}
\BIBentrySTDinterwordspacing

\bibitem{DBLP:journals/corr/KingmaB14}
\BIBentryALTinterwordspacing
D.~P. Kingma and J.~Ba, ``Adam: {A} method for stochastic optimization,'' in
  \emph{3rd International Conference on Learning Representations, {ICLR} 2015,
  Conference Track Proceedings}, 2015. [Online]. Available:
  \url{http://arxiv.org/abs/1412.6980}
\BIBentrySTDinterwordspacing

\bibitem{levenshtein1966binary}
V.~I. Levenshtein, ``Binary codes capable of correcting deletions, insertions,
  and reversals,'' in \emph{Soviet physics doklady}, vol.~10, no.~8, 1966, pp.
  707--710.

\bibitem{dat:github}
\BIBentryALTinterwordspacing
J.~Rand and A.~Miranskyy. (2021) Log generating tool. [Online]. Available:
  \url{https://github.com/WulffHunter/log_generator}
\BIBentrySTDinterwordspacing

\bibitem{code:zenodo}
\BIBentryALTinterwordspacing
------. (2021) Log parsing: generators. [Online]. Available:
  \url{https://doi.org/10.5281/zenodo.4536575}
\BIBentrySTDinterwordspacing

\bibitem{dat:zenodo}
\BIBentryALTinterwordspacing
------. (2021) Log parsing: datasets. [Online]. Available:
  \url{https://doi.org/10.5281/zenodo.4536514}
\BIBentrySTDinterwordspacing

\end{thebibliography}

\end{document}